%% file: article.tex
\documentstyle[psfig]{europhys}

\input euromacr.tex

\begin{document}

\euro{}{}{}{}

\Date{}

\shorttitle{J. BADRO \etal A COMBINED XAS AND XRD ETC.}

\title{\boldmath A combined XAS and XRD Study of the High-Pressure
Behaviour of $GaAsO_{4}$ Berlinite}

\author{James Badro\inst{1}, Philippe Gillet\inst{1}, Paul F. McMillan\inst{2},
Alain Polian\inst{3} \And Jean-Paul Iti\'{e}\inst{3}}

\institute{
     \inst{1} Laboratoire de Sciences de la Terre,
Ecole Normale Sup\'{e}rieure de Lyon\\
46, all\'{e}e d'Italie 69364 Lyon Cedex 07, France.\\
     \inst{2} Materials Research Group in High Pressure
Synthesis,  Department of Chemistry \\
and Biochemistry,  Arizona State University, Tempe, AZ 85287, USA.\\
     \inst{3} Physique des Milieux Condens\'{e}s,
Universit\'{e} Pierre et Marie Curie,  B77\\
 4,  place Jussieu 75252 Paris Cedex 05, France.
}

\rec{}{}


\maketitle

\begin{abstract}
Combined X-ray absorption spectroscopy (XAS) and X-ray diffraction (XRD)
experiments have been carried out on $\alpha -GaAsO_{4}$ (berlinite
structure) at high pressure and room temperature.  XAS measurements
indicate four-fold to six-fold coordination changes for both cations.
The two local coordination transformations occur at different rates but
appear to be coupled.  A reversible transition to a high pressure
crystalline form occurs around 8~GPa.  At a pressure of about 12~GPa,
the system mainly consists of octahedral gallium atoms and a mixture of
arsenic in four-fold and six-fold coordinations.  A second transition
to a highly disordered material with both cations in six-fold
coordination occurs at higher pressures and is irreversible.
\end{abstract}

\section{Introduction}
The mineral quartz ($\alpha -SiO_{2}$) and its ordered $AlPO_{4}$
(berlinite) analogue form archetypal tetrahedral framework structures
taken by many compounds ($GeO_{2}$,  $GaPO_{4}$,  $GaAsO_{4}$,
$AlAsO_{4}$) which exhibit phase transformations under high pressure.
At room temperature,  the kinetics of these pressure-induced
transitions are slow and metastable crystal-to-crystal (C-C)
transitions or pressure-induced amorphization (PIA) processes are observed
\cite{hemley1,hemley2,meade1,mcneil1,kingma1,silica}.  The room
temperature compression behaviour and PIA of $\alpha -GeO_{2}$ was
studied previously by X-ray diffraction and X-ray absorption
spectroscopy (XAS) \cite {itie1}.  The results showed that the
germanium coordination rises from four to six through the
high pressure transformation.  Molecular dynamics (MD) simulations
\cite{tse1,tse2,chelikowsky1,bing1,watson1,badro1,silica,silicatemelts}
on $SiO_{2}$ quartz indicate that the PIA is also associated with a
coordination change.
 
In the case of quartz-like berlinites,  it was first thought that
$\alpha -AlPO_{4}$ exhibited the same room temperature and high
pressure behaviour as quartz,  and underwent PIA at 15~GPa
\cite{kruger1}.  A more recent study \cite{gillet2} has shown that
under hydrostatic conditions,  this transition is not PIA,  but a
polymorphic crystalline phase transition.  The high pressure form is
poorly crystallized,  and has a very weak X-ray diffraction pattern.
It is suggested from MD simulation that the transition is associated
with destabilization of the $AlO_{4}$ tetrahedron relative to $
AlO_{6}$.  {\it In situ} XAS experiments were carried out on the
isoelectronic -isostructural compound,  $GaPO_{4}$,  at the K edge
absorption energy of gallium \cite{itie3,polian2}.  The polymorphic
phase transition observed at 13~GPa is associated with a four- to
six-fold oxygen coordination change observed around gallium atoms.  In
the present work,  we have used combined XAS and XRD methods to study
the high pressure behaviour of $GaAsO_{4}$,  and investigate the local
environment of both cations.

\section{Experimental}
Powdered $\alpha -GaAsO_{4}$ \cite{matar1} was used for these high
pressure energy dispersive EXAFS and X-ray diffraction studies.  The
samples were loaded in a stainless steel gasket with a 200 $\mu m$ hole
diameter and 50 $ \mu m$ initial thickness and high pressures were
generated by a membrane driven diamond anvil cell (DAC).  For
diffraction,  three independent experiments were carried out on the
DW-11 energy dispersive X-ray diffraction beamline on the DCI ring of
the LURE synchrotron facility in Orsay,  France, using argon and
silicone oil as pressure transmitting media, and ruby fluorescence as a
pressure gauge.  XAS was performed on the D11 dispersive XAS
\cite{dartyges1,tolentino1} beamline at both Ga and As K edges using
silicone oil as the pressure transmitting medium.

\section{Results : X-ray Diffraction}
\begin{figure}[h]
\psfig{figure=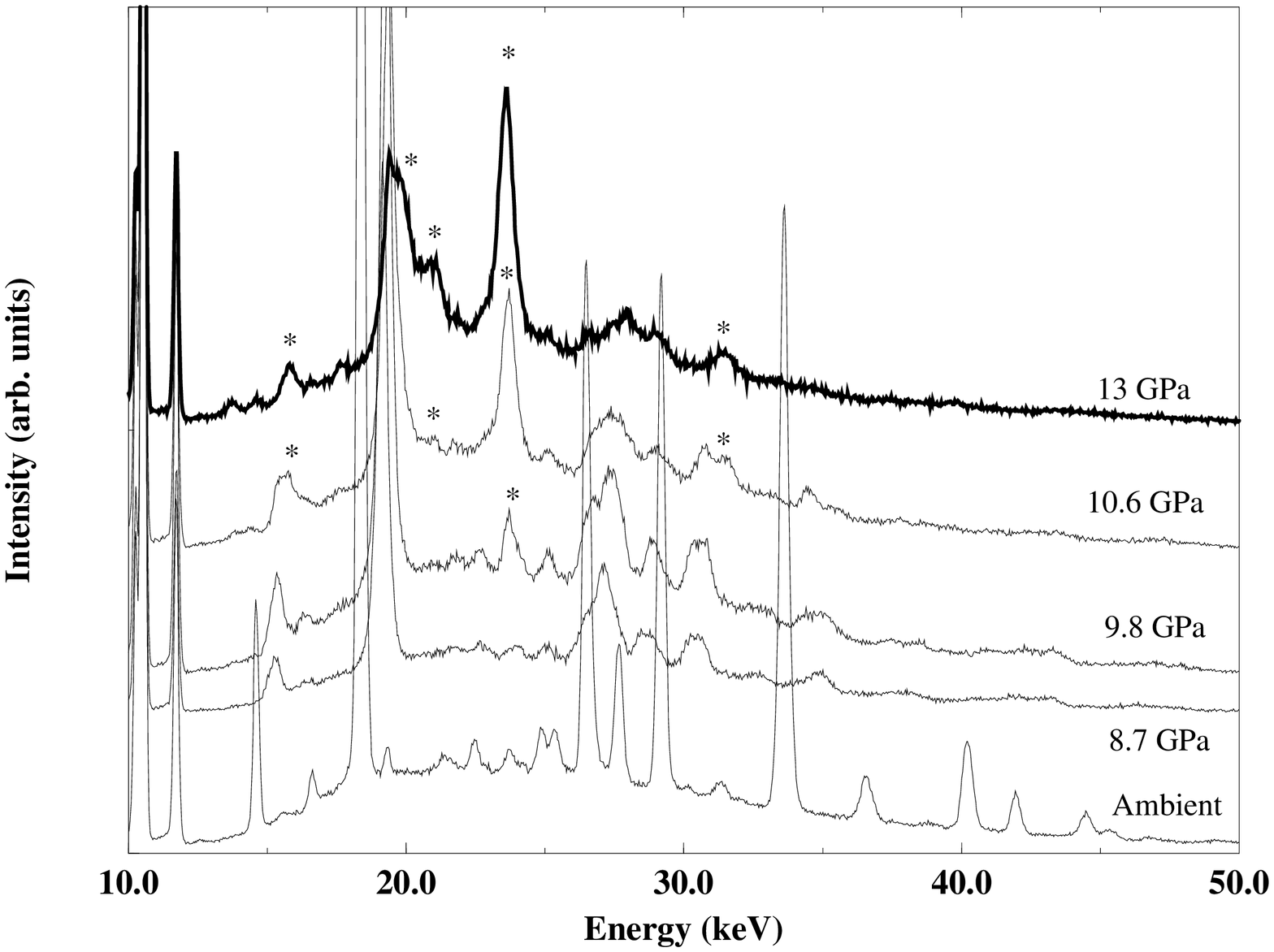,angle=270,width=5cm}
\vspace{-10mm}
\caption{Energy dispersive X-ray diffraction spectra for $GaAsO_{4}$ as
a function of pressure. One can note that a phase transition occurs
between 8.7 and 9.8~GPa.  A new line appears at 24 keV (*-sign),  but a
total breakdown of the hexagonal structure does not occur until above
10.6~GPa (boldened spectrum).  The diffraction angle is
$2\theta$=11.28$^\circ$.}
\label{xrd}
\end{figure}

\begin{figure}[h]
\psfig{figure=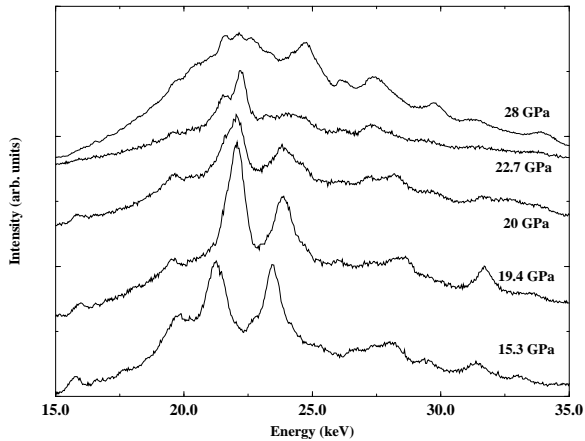,angle=270,width=5cm}
\vspace{-10mm}
\caption{Above 20~GPa, new diffraction lines appear in the spectrum as
the peaks of the intermediate phase broaden and weaken. They are
assigned to the 6-fold coordinated phase observed by EXAFS
spectroscopy. Once again, this phase seems to appear at lower pressures
but the weakness of its related diffraction lines is such that they are
not observed until the preceding intermediate structure's lines
weaken themselves.}
\label{xrdhp}
\end{figure}

\begin{figure}[h]
\centerline{\psfig{figure=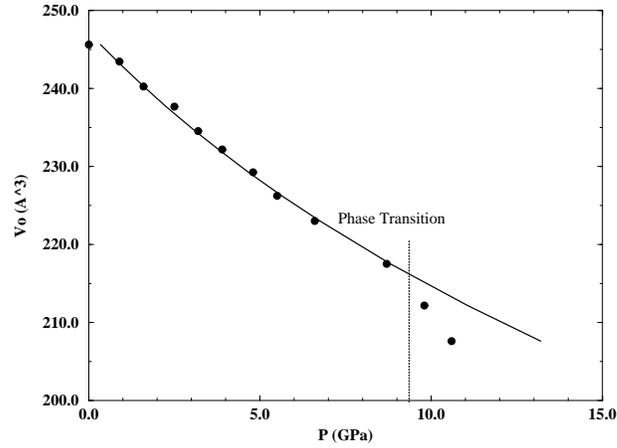,angle=270,width=5cm}}
\vspace{-10mm}
\caption{Pressure-Volume data points obtained from refinement of the
hexagonal structure of the sample in the low pressure phase.  The
points are fitted by a third order Birch-Murnaghan EOS with $K_{0}=53.4
\;GPa$, $K^{^{\prime}}_{0}$ fixed to 4 and $V_{0}=247.2 \;$ \AA $^{3}$
for the data points up to 8.7~GPa. Above this pressure, the structure
distorts and the system is a mixture of two phases thus rendering
volume data unreliable and giving yet another signature of the phase
transition.}
\label{eos}
\end{figure}

X-ray diffraction spectra recorded as a function of pressure are
reproduced in Fig.  \ref{xrd}.  A polymorphic crystal-to-crystal phase
transition is observed to begin \cite{itie3,clark1} around 8.7~GPa.  A
new diffraction peak is observed between 8.7 and 9.8~GPa,  and the
remaining peaks begin to broaden. Meanwile, a fit of the low pressure
phase's peaks using the original hexagonal cell yields a negative
pressure derivative of the bulk modulus ($K^{^{\prime }}$) above the
transition (fig. ~\ref{eos}), which is physically unreasonable and
characteristic of a phase mixture.  The other very weak and broad peaks
of the high pressure phase could be clearly observed only after the
hexagonal phase has vanished around 13~GPa. Further pressurization
results in a compression of this high pressure structure up to 22~GPa.
Above 22~GPa,  new peaks appear and superpose with the previous
spectrum, showing the onset of a second phase transition.  At 28~GPa,
the highest pressure reached in this EDX study, the spectrum is that of
a very poorly crystallized sample (Fig.  \ref{xrdhp}).

The isothermal equation of state (EOS) obtained from structure
refinements using the DICVOL structure refinement program \cite{dicvol}
in the low-pressure hexagonal phase (up to 8.7 GPa) and fitted by a
third order Birch-Murnaghan EOS shows a bulk modulus $K_{0}=53.4\;GPa$
with a pressure derivative fixed to $K_{0}^{^{\prime }}=4$ and a unit
cell volume at room pressure $V_{0}=247.2\;$\AA $^{3}$ (Fig.
~\ref{eos}).  This is consistent with previously reported volume data
\cite{matar1,astm},  but our equation of state differs considerably
from that of Clark {\it et al. } \cite{clark1} ($ K_{0}=18.6\;GPa$ and
$K_{0}^{^{\prime }}=18.8)$.

Given that no structural data could be extracted from the diffraction
patterns and that therefore no structural refinements coulb de carried
out for the two high pressure phases, we chose to combine XAS data in
order to analyse the local transformations occuring around both cations.

\section{Results : X-ray Absorption Spectroscopy}
\begin{figure}[h]
\centerline{\psfig{figure=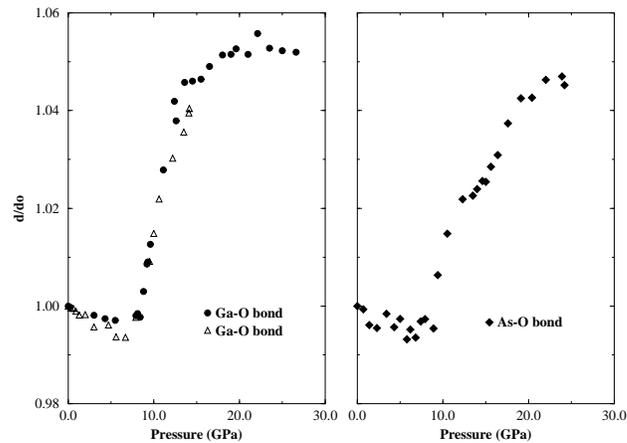,angle=270,width=5cm}}
\vspace{-10mm}
\caption{Relative bond length ($d/d_{0}$) for the Ga-O (two independent
experiments) and As-O bonds.  It can be seen that there is one
compression regime in the 4-to-6 transformation concerning the gallium
local environment,  but two for arsenic.  The second regime appears for
arsenic at the pressure for which gallium has become totally octahedral
(12-13~GPa).}
\label{exafs}
\end{figure}

X ray absorption spectroscopy measurements were performed on the system
both at the Ga and As K edges and the cation-oxygen distances deduced
from the experimental data using the CDXAS package \cite{sanmiguel1}
are shown in figure~\ref{exafs}; In the low-pressure phase,  the Ga-O
and As-O bond length moduli can be measured; they both lie around
$K_{0}^{bond}=320\;GPa$ with $K_{0}^{^{\prime }}$ set to 4.  The large
difference with the bulk modulus ($ K_{0}=53$ GPa) and the fact that
$K_{0}^{bond} \gg  3 K_{0}^{tet}$ shows that the compression mechanism
consists mainly of relative tilting of the tetrahedra (bending of the
T-O-T angle) as already reported by vibrational spectroscopy and MD
studies \cite{williams1,tse4}.
The onset of the polymorphic transition observed by X-ray diffraction
occurs around 8~GPa.  This corresponds to dramatic bond length changes
observed on both edges.  The Ga-O mean distance rises rapidly between 8
and 12.5~GPa, at which point the Ga-O bond length corresponds
approximately to that for octahedral $GaO_{6}$ groups.  This is also
indicated by the change in shape of the related XANES (X-ray absorption
near edge structure) spectra.
There is a further slight increase in Ga-O mean distance up to 22~GPa,
at which point the structural transformation to six-fold coordinated
gallium is complete.  Meanwhile,  the As-O mean distance also increases
between 8 and 12.5~GPa at which point another compression regime
appears and the rate of bond length change diminishes, until the
corresponding transition is complete at about 22~GPa.
These distance variations, {\it i.e.\/} the slope of the bond
length-pressure curves,  indicate the rate of transformation from
4-fold to 6-fold coordinated cations; gallium atoms undergo the
coordination change more rapidly with increasing pressure,  and
therefore at or about 12.5~GPa, the structure mainly consists of
six-fold coordinated gallium and arsenic atoms in an average
coordination intermediate between four and six.

\begin{figure}[h]
\centerline{\psfig{figure=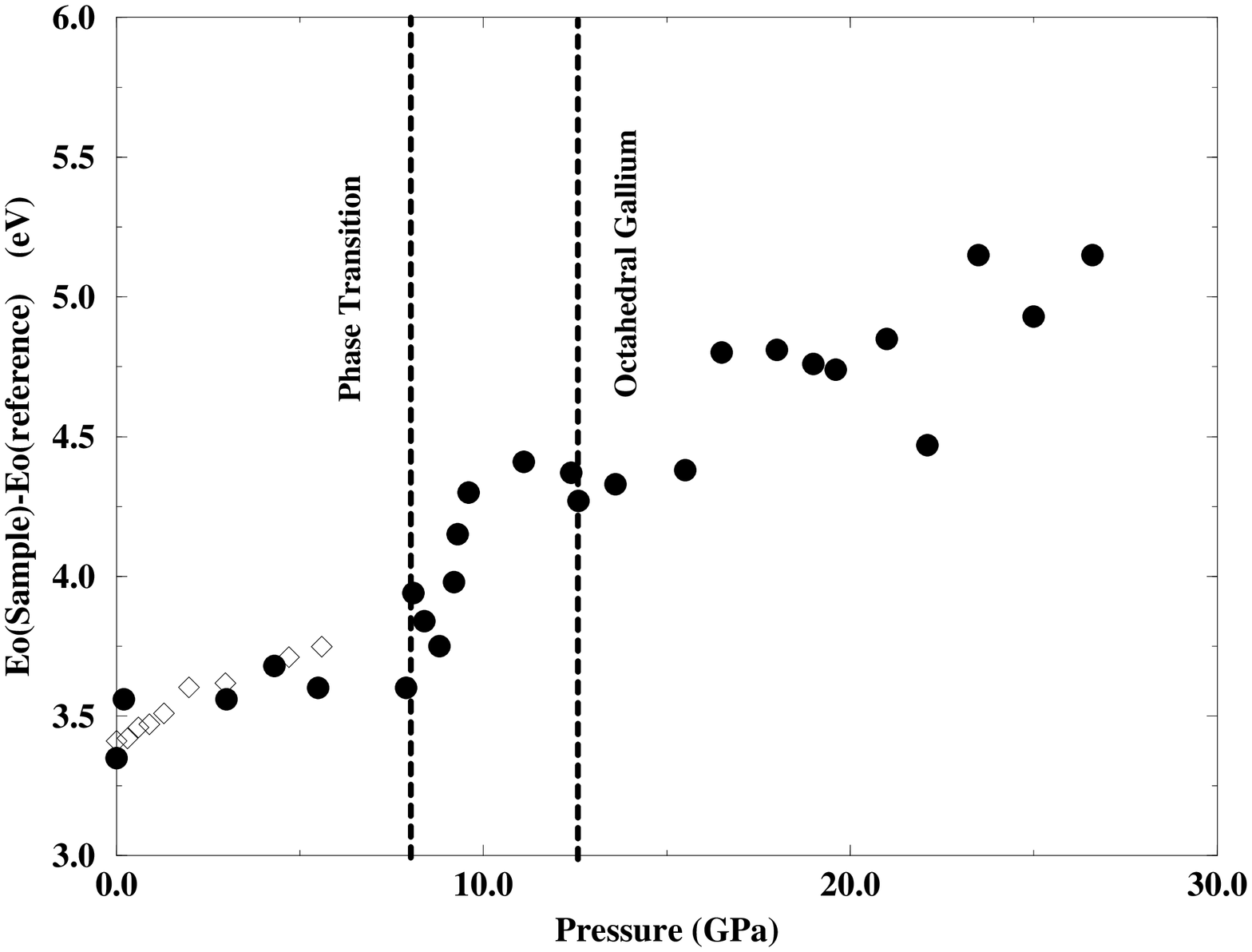,angle=270,width=5cm}}
\vspace{-10mm}
\caption{Gallium K edge energy displacement as a function of pressure.
On the ordinate axis, we report the difference between the K edge
absorption energy of gallium in the $GaAsO_4$ sample at pressure $P$
and that of the K edge absorption energy of gallium in a $GaAs$
reference crystal. The transition is accompanied by an increase in the
absorption energy.  This plot shows that there is notable change in the
core level electronic structure associated with the bonding change
throughout this transition.}
\label{edge}
\end{figure}

The gallium K edge energy as a function of pressure reported in
figure~\ref {edge}  shows that the transition to the octahedral
configuration is accompanied by a 30\% increase in the absorption
energy (with repect to the K edge absortion level of a $GaAS$
standard), and that therefore the different nature of the bonding
resulting from modifications of the uppermost electronic shell can be
detected by accurate K edge absorption energy measurements.

\section{Discussion : Compression}
It is interesting to note the correlation between XAS and XRD
measurements.  The cation--oxygen distance starts increasing around
8~GPa,  monitored by XAS.  Above this pressure,  the diffraction lines
of the sample start broadening, the background scattering intensity
rises,  a new diffraction line appears (fig.~\ref{xrd}) and the cell
parameters and volumes calculated by the data refinement for the
hexagonal structure behave unreasonably, in that sense that the mixture
of two phases renders the volume fit false, hence a rising difference
between the extrapolated EOS and the calculated volume of the hexagonal
phase.
Around 12-13~GPa, essentially all gallium atoms become 6-fold
coordinated, whereas only part of the arsenic is octahedral; At this
pressure, a dramatic change is observed by XRD, with the signature of a
disordered crystalline material (weak and broad lines,  intense broad
background).  At this pressure,  the hexagonal peaks and therefore the
associated phase have totally disappeared.

Given these combined observations,  we argue that the high pressure
phase appears at 8 GPa.  Between 8 and 13 GPa, the system is
mixed-phase with the average coordination intermediate between 4 and 6
and an average bond length intermediate between $l_{tet}$ and
$l_{oct}$.  In this pressure domain,  only a small change of the
diffraction spectra occurs,  because of the superposition of the weak
spectrum of the high pressure phase with the intense spectrum of the
hexagonal structure.  It was not possible to refine the new diffraction
peaks. The previously proposed distorted monoclinic structure
\cite{clark1} does not match our diffraction data.  Nevertheless it
appears from EXAFS that this intermediate phase has twice as much
octahedral gallium than arsenic, because the rate of four-fold to
six-fold coordination change with pressure is twice as large for
gallium than for arsenic (see fig.~\ref{exafs}), which seems to confirm
one of the structural observations of Clark {\it et al.}.
From our data,  we can confirm that the intermediate phase is not of
the $InPO_{4}$ type,  an expected intermediate structure in berlinite
transitions from quartz to rutile-like phases, but belongs to a lower
symmetry structure because of the 1/3--2/3 mixture of six-coordinated
arsenic and gallium atoms respectively.

Further increase in pressure above 13~GPa reveals another structural
transformation.  First, continued compression results in a slower rate
of change in the arsenic coordination, because the six-coordinated
structure of gallium atoms renders the system less compressible.  At
approximately 22~GPa, all arsenic clusters have transformed into
octahedra and the system consists of a pseudo-rutile type structure
with fully octahedral cation arrangements, which is indicated by the
XANES spectra at the As and Ga K edges which show a clear signature of
the well known octahedral clusters \cite{itie1}.
The weak diffraction patterns at these pressures do not permit a
structure refinement of this phase.  It is not easy to assign a
transition pressure,  but it is reasonable to suppose that this
pseudo-rutile phase appears when arsenic atoms enter their second
compression regime around 15 GPa,  just after the hexagonal phase has
vanished. Above 22 GPa,  the system contains the pseudo-rutile phase.

\section{Discussion : Decompression}
The latter assumption, concerning the onset pressure of the second
phase transition can be further justified by the decompression history.
In fact, one expects the intermediate phase to revert to the original
material upon decompression, as observed on berlinite phosphates,
whereas the second would only give rise to a partially or totally
amorphous sample on pressure quench.

Samples quenched from 15~GPa,  {\it i.e.\/} after the hexagonal phase
has vanished and the second phase has started appearing, undergoes a
back-transformation to the original structure (XRD analysis) on
decompression. The local atomic structure is tetrahedral but with a
partial amorphization of the sample as evidenced by the decrease of the
intensity of the XANES features (XAS analysis).  On the other hand, the
sample decompressed from 25~GPa is completely amorphous and its cations
remain mainly in sixfold coordination.
 
\section{Conclusion}
We have shown that gallium arsenate berlinite undergoes two phase
transitions at high pressure and 300 K.  The first transformation is a
crystalline polymorphic phase transition and is associated,  as opposed
to phosphates \cite {itie2,polian2,badro2,badro4},  with a local
transition from four- to six-fold coordination of {\it both} cations,
{\it i.e.\/} the high-pressure phase is not of the $InPO_{4}$ type.
Only part of the As atoms have transformed to high coordination at this
point.  Further compression leads to full transformation of both
cations to six-coordinate,  and the structure can be related to a
rutile-like phase.  As long as this phase has not appeared,  samples
decompressed allow recovery of the starting crystal.  The decompression
of systems consisting partly or entirely of this pseudo-rutile phase do
not return to the initial structure,  and a partly or entirely
amorphous material is recovered.

The destabilization of the tetrahedral clusters is confirmed by {\it in
situ} EXAFS. It appears that a large difference exists in the high
pressure behaviour of arsenate and phosphate berlinites. In the case of
arsenates, both cations undergo a transformation to sixfold
coordination at roughly the same pressure whereas the critical pressure
needed to destabilize the $PO_{4} $ tetrahedra is well above the first
transition point at 13--14~GPa for gallium or aluminium,  and the two
local cationic transitions are totally dissociated. The stability of
the $PO_{4}$ group is likely responsible for the total reversibility of
the phase transition for phosphate berlinites which is not observed for
arsenate berlinites (partial or complete amorphization on
decompression).  Further experiments under simultaneous high
pressure--high temperature conditions \cite{badro2,gillet3} are under
way to fully delineate the phase diagram.

\stars

Laboratoire des Sciences de la Terre is CNRS UMR 5570.  Physique des Milieux
Condens\'{e}s is CNRS URA 782.

\newpage

\end{document}

%% file: euromacr.tex
\def\etal{{\hbox{{\tenit\ et al.\/}\tenrm :\ }}}

\def\And{{\rm and\ }}

\def\stars{\bigskip\centerline{***}\medskip}

\newif\ifboo \boofalse

%% file: article.bbl
\begin{thebibliography}{10}

\bibitem{hemley1}
R.J. Hemley.
\newblock In {\it High-Pressure research in mineral physics}, Mineral Physics
  2. Terra Scientific Publishing Company -- AGU, 1987.

\bibitem{hemley2}
R.J. Hemley, A.P. Jephcoat, H.K. Mao, L.C. Ming, and M.H. Manghnani.
\newblock {\it Nature}, {\bf 334}:52--54, 1988.

\bibitem{meade1}
C.~Meade, R.J. Hemley, and H.K. Mao.
\newblock {\it Phys. Rev. Lett.}, {\bf 69}(9):1387--1390, 1992.

\bibitem{mcneil1}
L.E. McNiel and M.~Grimsditch.
\newblock {\it Phys. Rev. Lett.}, {\bf 68}:83--85, 1992.

\bibitem{kingma1}
K.J. Kingma, C.~Meade, R.J. Hemley, H.~Mao, and D.R. Veblen.
\newblock {\it Science}, {\bf 259}:666--669, 1993.

\bibitem{silica}
P.J. Heaney, C.T. Prewitt, and G.V. Gibbs, editors.
\newblock {\it Silica. Physical Behavior, Geochemistry and Materials
  Applications}, volume~29 of {\it Reviews in Mineralogy}.
\newblock Mineralogical Society of America, 1994.

\bibitem{itie1}
J.-P. Itié, A.~Polian, G.~Calas, J.~Petiau, A.~Fontaine, and H.~Tolentino.
\newblock {\it Phys. Rev. Lett.}, {\bf 63}(9):389--401, 1989.

\bibitem{tse1}
J.S. Tse and D.D. Klug.
\newblock {\it Phys. Rev. Lett.}, {\bf 67}(25):3559, 1991.

\bibitem{tse2}
J.S. Tse and D.D. Klug.
\newblock {\it Science}, {\bf 255}:1559--1561, 1992.

\bibitem{chelikowsky1}
N.~Bingelli, N.~Troullier, J.-L. Martins, and J.R. Chelikowsky.
\newblock {\it Phys. Rev. B}, {\bf 44}(2):4471, 1991.

\bibitem{bing1}
N.~Binggeli, N.R. Keskar, and J.R. Chelikowsky.
\newblock {\it Phys. Rev. B}, {\bf 49}:3075, 1994.

\bibitem{watson1}
G.W. Watson and S.C. Parker.
\newblock {\it Philosophical Mag. Lett.}, {\bf 71}(1):59--64, 1995.

\bibitem{badro1}
J.~Badro, J.-L. Barrat, and Ph. Gillet.
\newblock {\it Phys. Rev. Lett.}, {\bf 76}(5):772--775, 1996.

\bibitem{silicatemelts}
J.F. Stebbins, P.F. McMillan, and D.B. Dingwell, editors.
\newblock {\it Structure, Dynamics and Properties of Silicate Melts}, volume~32
  of {\it Reviews in Mineralogy}.
\newblock Mineralogical Society of America, 1995.

\bibitem{kruger1}
M.B. Kruger and R.~Jeanloz.
\newblock {\it Science}, {\bf 249}(647):647--649, 1990.

\bibitem{gillet2}
Ph. Gillet, J.~Badro, B.~Varrel, and P.F. McMillan.
\newblock {\it Phys. Rev. B}, {\bf 51}(17):11262--11269, 1995.

\bibitem{itie3}
J.-P. Itié, T.~Tinoco, A.~Polian, G.~Demazeau, S.~Matar, and E.~Philippot.
\newblock {\it High Pressure Research}, {\bf 14}:269--276, 1996.

\bibitem{polian2}
A.~Polian, J.-P. Itié, and J.~Badro.
\newblock {\it Unpublished}, {\bf }.

\bibitem{matar1}
S.~Matar, M.~Lelogeais, D.~Michau, and G.~Demazeau.
\newblock {\it Materials Letters}, {\bf 10}(1,2), 1990.

\bibitem{dartyges1}
E.~Dartyges, C.~Depautex, J.M. Dubuisson, A.~Fontaine, A.~Jucha, P.~Leboucher,
  and G.~Tourillon.
\newblock {\it Nucl. Inst. Meth.}, {\bf A246}:456, 1986.

\bibitem{tolentino1}
H.~Tolentino, E.~Dartyges, A.~Fontaine, and G.~Tourillon.
\newblock {\it J. Appl. Phys.}, {\bf 21}:15, 1988.

\bibitem{clark1}
S.M. Clark, A.G. Christy, R.~Jones, J.~Chen, J.M. Thomas, and G.N. Greaves.
\newblock {\it Phys. Rev. B}, {\bf 51}(1):38--51, 1995.

\bibitem{dicvol}
A.~Boultif and D.~Louer.
\newblock {\it J. Appl. Cryst.}, {\bf 24}:987--993, 1991.

\bibitem{astm}
{\it JCPDS}, {\bf 31-541}, 1981.

\bibitem{sanmiguel1}
A.~San Miguel.
\newblock {\it Physica B}, {\bf 208-209}:177, 1995.

\bibitem{williams1}
Q.~Williams and R.~Jeanloz.
\newblock {\it Nature}, {\bf 338}, 1989.

\bibitem{tse4}
J.S. Tse and D.D. Klug.
\newblock {\it Phys. Rev. Lett.}, {\bf 70}(2), 1993.

\bibitem{itie2}
J.-P. Itié, A.~Polian, D.~Martinez, V.~Briois, A.~DiCicco, A.Filipponi, and
  A.~San Miguel.
\newblock {\it J. Physique (Paris)}, {\bf }(Accepted), 1997.

\bibitem{badro2}
J.~Badro, Ph. Gillet, J.-P. Itié, and A.~Polian.
\newblock {\it Unpublished}, {\bf }.

\bibitem{badro4}
J.~Badro, J.-P. Itié, A.~Polian, and Ph. Gillet.
\newblock {\it J. Physique (Paris)}, {\bf }(Accepted), 1997.

\bibitem{gillet3}
Ph. Gillet, G.~Fiquet, I.~Daniel, and B.~Reynard.
\newblock {\it Geophys. Res. Lett.}, {\bf 20}:1931--1934, 1993.

\end{thebibliography}
